\def\etal{{\it et al.\thinspace}}
\def\mearth{{\rm\,M_\oplus}}
\def\msun{{\rm\,M_\odot}}

\documentclass[apjl]{emulateapj}

\accepted{Sept 18, 2008}

\lefthead{}
\righthead{}

\begin{document}

\shorttitle{Mean Motion Resonances from Planet-Planet Scattering}
\shortauthors{Raymond et al.}

\title{Mean motion resonances from planet-planet scattering}

\author{Sean N. Raymond\altaffilmark{1}, 
Rory Barnes\altaffilmark{2},
Philip J. Armitage\altaffilmark{3},
\& Noel Gorelick\altaffilmark{4}}

\altaffiltext{1}{NASA Postdoctoral Program Fellow, Center for Astrophysics and
Space Astronomy, University of Colorado, Boulder CO 80309; raymond@lasp.colorado.edu}
\altaffiltext{2}{Lunar and Planetary Laboratory, University of Arizona,
Tucson, AZ; rory@lpl.arizona.edu }
\altaffiltext{3}{JILA, University of Colorado, Boulder CO 80309; pja@jilau1.colorado.edu}
\altaffiltext{4}{Google, Inc., 1600 Amphitheatre Parkway, Mountain View, CA
94043; gorelick@google.com}

\begin{abstract}
Planet-planet scattering is the leading mechanism to explain the large
eccentricities of the observed exoplanet population.  However, scattering has
not been considered important to the production of pairs of planets in mean
motion resonances (MMRs).  We present results from a large number of numerical
simulations of dynamical instabilities in 3-planet systems.  We show that MMRs
arise naturally in about five percent of cases.  The most common resonances we
populate are the 2:1 and 3:1 MMRs, although a wide variety of MMRs can occur,
including high-order MMRs (up to eleventh order).  MMRs are generated
preferentially in systems with uneven mass distributions: the smallest planet
is typically ejected after a series of close encounters, leaving the
remaining, more massive planets in resonance.  The distribution of resonant
planets is consistent with the phase-space density of resonant orbits, meaning
that planets are randomly thrown into MMRs rather than being slowly pulled
into them.  It may be possible to distinguish between MMRs created by
scattering vs. convergent migration in a gaseous disk by considering planetary
mass ratios: resonant pairs of planets beyond $\sim$ 1 AU with more massive
outer planets are likely to have formed by scattering.  In addition,
scattering may be responsible for pairs of planets in high-order MMRs (3:1 and
higher) that are not easily populated by migration.  The frequency of MMRs
from scattering is comparable to the expected survival rate of MMRs in
turbulent disks.  Thus, planet-planet scattering is likely to be a major
contributor to the population of resonant planets.
\end{abstract}


\section{Introduction}

The current sample of exoplanets exhibits several interesting dynamical
features (Butler \etal 2006): here we focus on two of these.  First, there is
a vast range of planetary eccentricities, from zero to $>0.9$, with a median
of 0.2 (0.27 for planets past 0.1 AU that have not been affected by tides;
Rasio \etal 1996; Jackson \etal 2008).  Second, mean motion resonances (MMRs)
in multiple planet systems appear to be relatively common (e.g., Marcy
\etal 2001).  There are 31 currently-known multiple planet systems comprising
44 pairs of adjacent planets, of which ten (23\%) show some evidence of
resonances (Table 1).  However, the evidence for resonances is tentative for
all but a few cases.

Dynamical instabilities in systems of two or more planets can explain the wide
eccentricity distribution of exoplanets (Rasio \& Ford 1996; Weidenschilling
\& Marzari 1996; Lin \& Ida 1997).  Instabilities arise on timescales that are
related to the planets' initial separation (Marzari \& Weidenschilling 2002;
Chatterjee \etal 2008), and lead to close encounters between planets and
subsequent ejections or mergers.  In the aftermath of close encounters, the
surviving planets can statistically reproduce the observed eccentricity
distribution of exoplanets (Adams \& Laughlin 2003; Juric \& Tremaine 2008;
Chatterjee \etal 2008).

MMRs are thought to arise primarily from convergent migration in gaseous
protoplanetary disks (Snellgrove \etal 2001; Lee \& Peale 2002).  Indeed,
models show that capture in the 2:1 and 3:2 MMRs is a particularly common
occurrence (Thommes 2005; Pierens \& Nelson 2008; Lee \etal 2008).  However,
MRI-derived turbulence can act to remove planets from resonance.  Adams \etal
(2008) estimate that only 1\% of resonant systems should remain for a disk
lifetime of 1 Myr.

In this paper we attempt to reconcile the planet-planet scattering scenario
with the population of resonant exoplanets.  We numerically
investigate dynamical instabilities in systems of three planets located at
$\sim$ 2-10 AU with a variety of mass distributions.  We find that MMRs are a
common occurrence, arising in 5-10 percent of unstable systems.  Our
simulations populate a range of MMRs, including the 2:1, 3:2, 3:1, 4:1 and
extending up to much higher-order (Table 2).  MMRs are populated by scattering
at random into stable regions; the density of resonant orbits (i.e., the
fraction of phase space that undergoes resonant oscillations) is consistent
with the scattered resonant systems.  We propose several ways to discriminate
between scattering and convergent migration as the source of exoplanet MMRs. 

\section{Methods}

Our simulations started with three planets randomly separated by 4-5 mutual
Hill radii ($R_{H,m} = 0.5 (a_1+a_2) ([M_1+M_2]/3M_\star)^{1/3}$, where $a$ is
the semimajor axis and $M$ the mass).  This spacing was chosen to produce
instabilities on timescales of at least the $\sim 10^5$ year timescale of
runaway gas accretion\footnote{Indeed, instabilities occurred on timescales
from 100 years to 98 Myr with a median of a few $\times 10^5$ years.  In
addition, about 1/4 of simulations were stable for 100 Myr which shows that we
started close to the stability boundary.} (Pollack \etal 1996; Marzari
\& Weidenschilling 2002).  The outermost planet was placed two Hill radii
interior to 10 AU; cases with more massive planets and therefore larger Hill
radii therefore had the innermost planet closer to the star than for cases
with lower-mass planets (see below).  Planets were given zero eccentricity and
mutual inclinations of less than 1 degree, and the stellar mass was 1 $\msun$.

We considered a range of planetary mass distributions.  For our two largest
sets (1000 simulations each) we randomly selected planet masses according to
the observed distribution of exoplanet masses: $dN/dM \propto M^{-1.1}$
(Butler \etal 2006).  In the ``mixed1'' set we restricted the planet mass
$M_p$ to be between a Saturn mass $M_{Sat}$ and three Jupiter masses
$M_{Jup}$.  For our ``mixed2'' set, the minimum planet mass was decreased to
10 $\mearth$.  We also performed four ``Mequal'' sets (500 simulations each)
with equal mass planets for $M_p = 30 \mearth$, $M_{Sat}$, $M_{Jup}$, and $3
M_{Jup}$. Finally, the ``Mgrad'' sets (250 simulations each) contained radial
gradients in $M_p$.  For the JSN set, in order of increasing orbital distance,
$M_p$ = $M_{Jup}$, $M_{Sat}$, and $30 \mearth$.  For the NSJ set, these masses
were reversed, i.e., the $M_{Jup}$ planet was the most distant.  The 3JJS and
SJ3J sets had, in increasing radial distance, $M_p$ = $3 M_{Jup}$, $M_{Jup}$
and $M_{Sat}$, and $M_p$ = $M_{Sat}$, $M_{Jup}$ and $3 M_{Jup}$, respectively.

\begin{figure}[t]
\centerline{\plotone{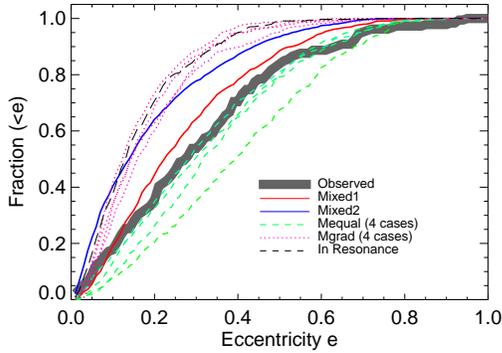}}
\caption{Cumulative eccentricity distribution of the known exoplanets
beyond 0.1 AU (thick grey), compared with our scattering simulations.}
\label{fig:edist}
\end{figure}

Each simulation was integrated with the hybrid version of the {\tt Mercury}
integrator (Chambers 1999).  All planets were assigned physical densities of
1.3 $g \,cm^{-3}$ and collisions were treated as inelastic mergers.  We used a
20 day timestep which tests show introduces an error of less than 1 part in
10$^5$ for perihelion distances larger than 0.5 AU.  In almost all cases
energy was conserved to better than one part in 10$^4$ for the entire 100 Myr
simulation, which Barnes \& Quinn (2004) showed is adequate precision to test
stability.  However, in some cases, energy was poorly conserved; those cases
were rerun with a 5 day timestep.  After this step, simulations with poor
energy conservation were removed from the analysis.

\section{Results}

Figure~\ref{fig:edist} shows that four of our sets of simulations match the
exoplanet eccentricity distribution -- mixed1, Mequal:Jup, Mequal:Sat, and
Mequal:30 $\mearth$ -- with P values from K-S tests greater than 0.01.
However, given the increasing number of low-mass exoplanets, we believe that
our mixed1 and mixed2 simulations are the most realistic initial conditions.
If scattering is the source of exoplanet eccentricities, then
soon-to-be-discovered systems with lower-mass planets should indeed tend to
have lower eccentricities (Ford \& Rasio 2008).

\begin{figure}[t]
\centerline{\plotone{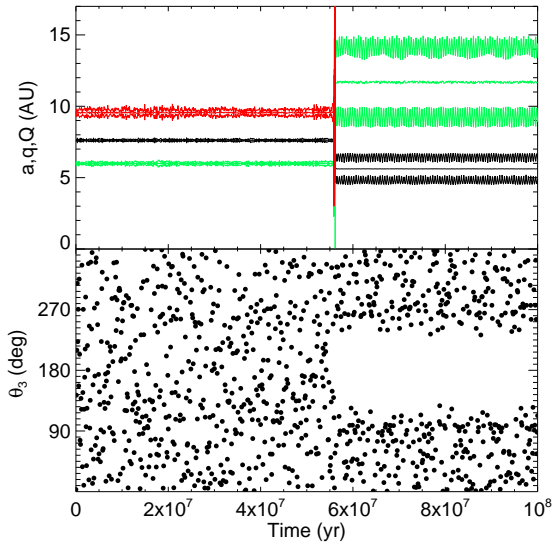}}
\caption{Evolution of a system that produced a pair of planets in the 3:1 MMR.
{\bf Top:} The three planets' semimajor axes $a$, perihelia $q$ and aphelia
$Q$.  The inner (green), middle (black), and outer (red) planets are 43, 105,
and 16 $\mearth$, respectively.  {\bf Bottom:} Evolution of the 3:1 resonant
argument $\theta_3 = 3 \lambda_2 - \lambda_1 - (\varpi_1+\varpi_2)$.  Resonant
libration starts immediately after ejection of the outer planet.}
\label{fig:evol31}
\end{figure}

We found MMRs by examining resonant arguments for simulations which produced
pairs of planets with period ratios close to commensurate values.  A pair of
planets is in resonance if any resonant argument $\theta_i$ librates rather
than circulates.  For MMR p+q : p, arguments are of the form
\begin{equation}
\theta_{i} = (p+q) \lambda_1 - p \lambda_2 -q \varpi_{1,2} \\
\end{equation}
\noindent where $\lambda$ are mean longitudes, $\varpi$ are longitudes
of pericenter, and subscripts 1 and 2 refer to the inner and outer planet,
respectively (e.g., Murray \& Dermott 1999).

Figure~\ref{fig:evol31} shows the evolution of a typical simulation that
created a resonant system.  The instability started 55.8 Myr into the
simulation, causing a series of close encounters. Within a few hundred
thousand years the outer planet was ejected and the inner two planets swapped
places.  The two remaining planets are on stable orbits in the 3:1 MMR, and
all three resonant arguments $\theta_{1,2,3}$ librate with amplitudes between
120$^\circ$ and 160$^\circ$.

A variety of MMRs is populated by scattering (see Table 2). Most common are
the 2:1 and 3:1, but higher-order MMRs exist up to eleventh order (13:2).  The
resonant libration amplitudes tend to be large, with a median of 110$^\circ$
and several cases with amplitudes of $\sim 170^\circ$.  This contrasts with
MMRs from migration which are created in a dissipative environment and should
be much smaller.  MMRs occur preferentially in cases with mixed mass
distributions, especially those with a positive mass gradient such as
Mgrad:NSJ.  MMRs are relatively rare for equal mass planets, and they tend to
arise more often after collisions rather than ejections, which contrasts with
the mixed and Mgrad cases.  This may explain why MMRs have not been found in
previous studies (except for isolated cases in Adams \& Laughlin 2003 and
Chatterjee \etal 2008).

MMRs appear to be populated at random: any stable region of parameter space
can be accessed by scattering.  To test this hypothesis, we calculated the
phase space density of resonant orbits within 10\% of the 2:1 and 3:1 MMRs for
planetary mass ratio of 1/3, 1, and 3, with $M_{inner}+M_{outer}=400 \mearth$.
For each MMR we ran $\sim$ 22,000 3-body (star + two planets) simulations for
1 Myr.  The semimajor axis of the inner planet was fixed at 5 AU (2:1 MMR) or
4 AU (3:1 MMR).  We sampled four parameters: the orbital period ratio, the
inner and outer planets' eccentricities, and the relative apsidal orientation.
Inclinations (of $<1^\circ$) and mean longitudes were sampled at random.
Resonant orbits were found by libration of resonant arguments.

\begin{figure}[t]
\centerline{\plotone{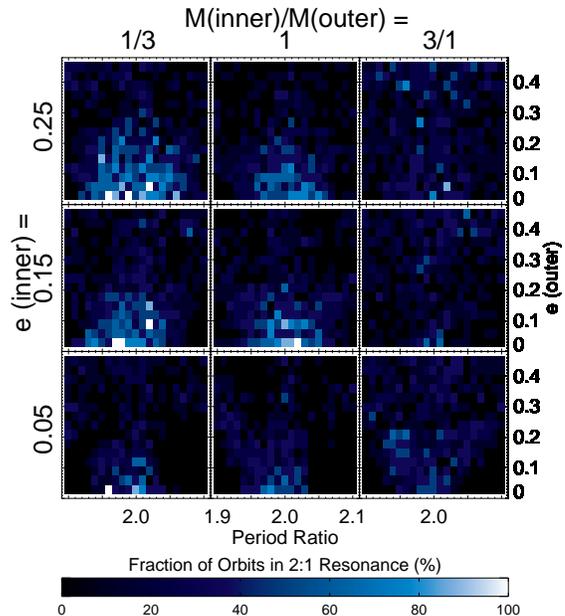}}
\caption{Resonant density for the 2:1 MMR.  For these simulations, the inner
planet's semimajor axis was fixed at 5 AU, and its eccentricity was varied
between 0.05 and 0.25.  The color corresponds to the fraction of orbits in
resonance for a given value of the outer planet's semimajor axis and
eccentricity (see color bar), averaged over 8 simulations with different
apsidal alignments.}
\label{fig:res21}
\end{figure}

The density of 2:1 resonant orbits is higher for a more massive outer planet
(Figure~\ref{fig:res21}).\footnote{For a more detailed study of the 2:1 MMR,
see Marzari \etal (2006) and Michtchenko \etal (2008a, 2008b).}  Almost all of
the 2:1 and 3:1 resonant orbits from scattering are found in areas of high
resonant density, and near-resonant ``false alarms'' lie in areas of low
density.  Thus, scattering does indeed appear to populate MMRs at random.
This explains why the 2:1 MMRs from our mixed1 and mixed2 simulations, with no
initial mass gradients, have a median $M_{inner}/M_{outer}$ of 0.5.  In
addition, the Mgrad:NSJ ($M_{inner}/M_{outer}\approx 1/3$) systems formed a
large number of 2:1 MMRs while the Mgrad:JSN ($M_{inner}/M_{outer} \approx 3$)
cases formed far fewer.  

For the parameter space we sampled, the integrated 3:1 resonant density is
$\sim 40\%$ less than the 2:1 density.  However, the available parameter space
is not evenly populated by scattering.  For example, scattering causes more
massive planets to be closer to the star (Chatterjee \etal 2008).  Indeed, the
integrated 2:1 resonant density for $M_{inner}/M_{outer} = 1/3$ is 66\% higher
than the 3:1 density.  This explains the increased number of 2:1 vs. 3:1 MMRs
in our sample -- 74 cases of the 2:1 MMR and 47 of the 3:1 (61\% more in 2:1).

\begin{deluxetable}{c|c|c|c|c}[t]
\scriptsize
\tablewidth{0pc}
\tablecaption{Candidate Resonant Planetary systems\tablenotemark{1}}
\renewcommand{\arraystretch}{.6}
\tablehead{
\\
\colhead{System} & 
\colhead{$a_1,a_2$} &
\colhead{$e_1,e_2$} &
\colhead{$M_1,M_2$} &
\colhead{MMR} \\ 
\colhead{(pair)} &
\colhead{(AU)} &
\colhead{ } &
\colhead{($M_{Jup}$)} }
\startdata
\\
GJ 876 c-b & 0.13, 0.2078 & 0.27, 0.025 & 0.56, 1.935 & 2:1\\

HD 73526 b-c & 0.66, 1.05 & 0.19, 0.14 & 2.9, 2.5 & 2:1\\

HD 82943 c-b & 0.746, 1.19 & 0.359, 0.219 & 2.01, 1.75 & 2:1\\

HD 128311 b-c & 1.099, 1.76 & 0.25, 0.17 & 2.18, 3.21 & 2:1\\

$\mu$ Arae d-b & 0.921, 1.497 & 0.067, 0.128 & 0.522, 1.676 & 2:1\\ 

GJ 317 b-c & 0.95, 2.35 & 0.19, 0.42 & 1.2, 0.83 & 4:1\tablenotemark{2} \\

HD 108874 b-c & 1.051, 2.68 & 0.07, 0.25 & 1.36, 1.018 & 4:1\\

HD 17156 b-c & 0.159, 0.481 & 0.6717, 0.136 & 3.111, 0.063 & 5:1\\

HD 202206 b-c & 0.83, 2.55 & 0.435, 0.267 & 17.4, 2.44 & 5:1\\
 
HD 208487 b-c & 0.49, 1.8 & 0.32, 0.19 & 0.45, 0.46 & 7:1\\

\enddata
\tablenotetext{1}{See http://www.lpl.arizona.edu/$\sim$rory/research/xsp/dynamics/.}
\tablenotetext{2}{Johnson \etal (2007) did not determine apsidal angles, but
Barnes \& Greenberg (2008) used a stability analysis to predict that the
system must be in the 4:1 MMR.}
\end{deluxetable}

The density of resonant orbits can explain other features of the population of
resonant planets.  The resonant planets tend to have smaller eccentricities
than the non-resonant planets (Fig.~\ref{fig:edist}).  Indeed, resonant
systems underwent an average of about five times fewer close encounters
between planets before the system stabilized ($\sim$ 30-50 vs. 100-$>$200
encounters), compared with the median outcome.  The time between the first
instability and stabilization for resonant cases was $\sim$ 50,000 years, a
factor of about five shorter than for the non-resonant systems.

Low-order MMRs preferentially arise in systems containing a planet with a
large Safronov number $S$.  $S$ is the ratio of the escape speed from a
planet's surface to the escape speed from the system, $S =
(M_p/M_\star)^{1/2}\, (a_p/R_p)^{1/2}$, where $M_\star$ is the stellar mass,
$a_p$ and $R_p$ are the planet's orbital distance and radius, respectively
(Safronov 1969).  Planets with larger $S$ give stronger velocity kicks and
thereby reduce the number of encounters needed to eject a planet.  Indeed, the
2:1 and 3:1 MMRs correlate with systems with at least one planetary $S$ value
above 4.  In the mixed2 set, which is the only set with a significant range in
$S$, the median $S_{max}$ for the 2:1 and 3:1 MMRs is 4.5, as compared with
3.5 for all unstable mixed2 simulations.  A K-S test shows that the two
samples are indeed different at the 99.9\% confidence level.  This constrains
where the 2:1 or 3:1 MMRs can arise as a function of $M_p$, $M_\star$, $a_p$,
and $R_p$: only systems with relatively high-mass planets ($M_p \gtrsim
M_{Jup}$) can generate these MMRs close-in.  In contrast, high-order (4:1 and
higher) MMRs from the mixed2 set match the sample of unstable cases and so are
not constrained.

The MMRs we found are numerically robust.  The median fractional integration
error dE/E for the 170 resonant systems was $2.6 \times 10^{-8}$, far below
the $\sim10^{-4}$ limit for determining stability (Barnes \& Quinn 2004), and
smaller than the median dE/E of $1.2 \times 10^{-7}$ for all unstable systems.
MMRs tend to arise in cases with short encounter times and relatively low
final eccentricities.  Those situations yield smaller dE/E than for the more
common stronger encounters that lead to very eccentric planets.  Thus, our
cutoff of dE/E $< 10^{-4}$ allows us to accurately sample both eccentricities
and MMRs.

\section{Discussion and Conclusions}

Planet-planet scattering creates MMRs.  The typical path to a resonant system
involves several close encounters between one smaller and two larger planets.
After a relatively short time of instability, the smaller planet is destroyed,
usually via ejection (78\% of all cases) or collision (22\%), leaving behind a
pair of resonant planets.  Relatively weak instabilities are probably very
common in planetary systems; one may even have occurred in our own Solar
System (Thommes \etal 1999).  Nonetheless, only a fraction of unstable systems
produce resonant planets, typically 5-10\%.  These systems have large
libration amplitudes and occupy a range of low-and high-order MMRs (Table 2).
Most of these resonances are indefinitely stable; we integrated the 170
resonant systems for an additional 1 Gyr and only 9 (5\%) left the resonance,
4 cases leading to an additional system instability.

\begin{deluxetable}{c|c|c|p{2cm}}[t]
\scriptsize
\tablewidth{0pt}
\tablecaption{Resonances from scattering simulations}
\renewcommand{\arraystretch}{.6}
\tablehead{
\\
\colhead{Set} & 
\colhead{Nsims ---} &
\colhead{N(\%) in} &
\colhead{MMRs}\\
\colhead{ } &
\colhead{unstable(\%)} &
\colhead{MMRs} &
\colhead{(\%)}}
\startdata
Mixed1 & 965--569 (59\%) & 27 (4.7\%) & 2:1 (1.6\%), 3:1 (1.6\%), 4:1
(0.7\%), 5:1, 6:1, 7:2, 9:2\\
Mixed2 & 982--744 (76\%) & 52 (7\%) & 3:2 (0.8\%), 2:1 (2.4\%), 3:1 (1.1\%),
5:3, 4:1, 5:2, 5:1, 7:3, 6:1, 7:2, 8:3, 9:4, 10:3, 12:5, 11:2, 14:5\\
Mequal:3$M_J$ & 368--241 (65\%) & 1 (0.4\%) & 7:1\\ 
Mequal:$M_J$ & 452--232 (51\%) & 4 (1.7\%) & 2:1, 3:1, 4:1 (0.9\%)\\
Mequal:$M_{Sat}$ & 390--362 (93\%) & 14 (3.9\%) & 2:1 (1.9\%), 3:1 (0.6\%),
5:2, 5:1 (0.6\%), 6:1, 7:1\\
Mequal:30$\mearth$ & 367--365 (99\%) & 10 (2.7\%) & 2:1 (1.9\%), 3:1, 11:6\\
Mgrad:JSN & 250--206 (82\%) & 13 (6.3\%) & 2:1 (1\%), 3:1 (1.9\%), 4:1,
5:2, 5:1, 7:3, 8:3, 13:2\\
Mgrad:NSJ & 245--221 (90\%) & 30 (14.6\%) & 2:1 (9.2\%), 3:1 (2.3\%), 5:2,
7:3, 7:2, 11:5\\
Mgrad:3JJS & 250--150 (60\%) & 4 (2.7\%) & 3:1, 4:1 (1.3\%), 11:3\\
Mgrad:SJ3J & 245--219 (89\%) & 16 (6.5\%) & 3:1 (5.7\%), 4:1
\enddata
\end{deluxetable}

It may be possible to tell apart resonant exoplanets created via scattering
from those created via convergent migration.  In fact, only one resonant
system appears to be inconsistent with a scattering origin due to its very
low-amplitude libration (GJ 876; Marcy \etal 2001).  If two planets are
trapped in the 2:1 or 3:2 MMR and the inner planet is the more massive, then
migration can be stopped or even reversed (Masset \& Snellgrove 2001; Crida \&
Morbidelli 2007).  However, if the outer planet is the more massive then
inward migration continues.  In contrast, scattering produces planets in a
variety of MMRs (including the 3:2 and 2:1) with a wide range in mass ratios
and a preference for the outer planet to be more massive.  Thus, scattering is
likely to be responsible for systems past $\sim$ 1 AU with 2:1 or 3:2 resonant
planets and a more massive outer planet.  The HD 128311 and $\mu$ Arae systems
are good candidates for creation via scattering (Table 1; see also S\'andor \&
Kley 2006).

Several extra-solar systems show tentative evidence for high-order MMRs --
4:1, 5:1 and even 7:1 (Table 1; Johnson \etal 2007; Correia \etal 2005;
Gregory 2007; Short \etal 2008).  No study to date has shown that migration
could capture planets in MMRs of higher order than 2:1, although we encourage
expanded studies of this process.\footnote{Highly-damped bodies can undergo
resonant shepherding by the 6:1 or even 8:1 MMRs (Raymond
\etal 2006; Mandell \etal 2007).  However, as bodies grow the damping
decreases, and shepherded planets do not survive in resonance.}  Scattering
produces a wide range of high-order resonances (Table 2).  Thus, if the
current candidate high-order MMR systems are confirmed (Table 1), then
scattering is likely to be the responsible mechanism.

Turbulence in gaseous disks may destroy MMRs, leaving perhaps only $\sim$ 1\%
of planet pairs in resonance (Adams \etal 2008).  This effect is stronger for
higher-order MMRs.  However, the timescale for MMR destruction is sensitive to
the strength of MRI turbulence which is very uncertain. In particular, if the
MRI is fully or partially suppressed by the low ionization fraction in the
inner protoplanetary disk (Gammie 1996) then the survival prospects for
resonant planets would be improved.  Nonetheless, if only a few percent of
systems remain in resonance, then scattering and migration may provide a
comparable number of MMRs.

Our simulations do not account for any dissipation.  However, instabilities
may occur while some gas remains in the disk (Moeckel \etal 2008).  In that
case, MMRs could still result from scattering (Lee \etal 2008) and perhaps
have smaller libration amplitudes.

In conclusion, we have identified a new mechanism for the creation of
exoplanet systems in MMRs.  Unfortunately the current data do not allow a
conclusive determination of a resonance, let alone precise descriptions of the
resonant argument oscillation.  Nonetheless, our scattering model has
important distinctions from the convergent migration model: high-order MMRs,
large-amplitude resonant libration, and low-order MMRs with
$M_{inner}/M_{outer} < 1$.  As the orbital properties of exoplanets are better
determined, it should be possible to distinguish between these scenarios.
 
\vskip .2in
We thank Google for access to their machines.  We are grateful to Greg
Laughlin, Dimitri Veras, and an anonymous referee for helpful input.
S.N.R. was supported by the NASA Postdoctoral Program administered by Oak
Ridge Associated Universities through a contract with NASA.  R.B. acknowledges
support from NASA's PG\&G grant NNG05GH65G and NASA Terrestrial Planet Finder
Foundation Science grant 811073.02.07.01.15.


\end{document}